# Telecom wavelength single-photon emission from quasi-resonantly excited InGaSb/AlGaSb quantum dots

TEEMU HAKKARAINEN,[1,2*] JOONAS HILSKA,[1] ARTTU HIETALAHTI,[3] SANNA RANTA,[1] MARKUS PEIL,[1] ROBERT MATYSIAK,[4] EMMI KANTOLA,[3] ABHIROOP CHELLU,[1] EFSANE SEN,[1] JUSSI-PEKKA PENTTINEN,[3] ANNA MUSIAŁ,[5] MICHAŁ GAWEŁCZYK,[4] MIRCEA GUINA[1]

*[1] Optoelectronics Research Centre, Physics Unit, Tampere University, FI-33720 Tampere, Finland*
*[2] Institute for Advanced Study, Tampere University, FI-33100 Tampere Finland*
*[3] VEXLUM Oy, FI-33710 Tampere, Finland*
*[4] Institute of Theoretical Physics, Wrocław University of Science and Technology, 50-370 Wrocław, Poland*
*[5] Department of Experimental Physics, Wrocław University of Science and Technology, 50-370 Wrocław, Poland*
**teemu.hakkarainen@tuni.fi*

**Abstract:** Deterministic light sources capable of generating quantum states on-demand at wavelengths compatible with fiber optics and atmospheric transmission windows are essential for practical applications in quantum communication, distributed photonic quantum computing, and quantum metrology. Currently, the technology providing semiconductor quantum emitters with the most promising properties is based on filling droplet-etched nanoholes to form quantum dots (QDs). However, the standard GaAs/AlGaAs material system does not offer telecom window emission. Here, we combine this growth method with antimonide-based materials to demonstrate single-photon emission at 1500 nm from a droplet-etched InGaSb QD. Our device with an antimony-based high refractive index contrast back-reflector designed for cryogenic operation and a solid immersion lens improves photon extraction. QD states are protected by a potential barrier limiting the influx of surrounding carriers, which however prevents revealing excitonic fine structure under nonresonant excitation. In this work, we employ a frequency-tunable continuous wave laser to achieve longitudinal optical (LO) phonon-assisted excitation of the QD ground state and resonant excitation of an excited state. These direct approaches for exciting a single InGaSb QD unlock access to its excitonic fine structure. The typical neutral biexciton-exciton cascade exhibits a negative binding energy of 1.4 meV (2.6 nm) and a fine structure splitting of 24.1±0.4 μeV. Furthermore, we obtain spectrally isolated emission from a charged exciton with a multi-photon probability of 5 % with LO phonon-assisted two-color excitation. These results represent a major step towards using this novel antimonide-based QD emitters as deterministic quantum light sources in complex quantum secure networks exploiting the wavelength compatibility with standard telecom fibers.

## 1. Introduction

Semiconductor light sources capable of generating photonic quantum states on demand are essential building blocks for applications like quantum communication,[1,2,3,4] quantum computing[5,6,7] and quantum metrology.[8] In particular, discrete-variable quantum cryptography[9] using light sources emitting single photons[10,11,12] or pairs of entangled photons[13,14] enables quantum-secure communication by means of quantum key distribution (QKD). QKD has already been demonstrated at system level in the BB84 scheme using InGaAs/GaAs quantum dots (QDs)[15] emitting at 920 nm and in the Ekert91 scheme using GaAs/AlGaAs QDs emitting polarization-entangled photons at 780 nm.[16] However, these pioneering demonstrations showcasing the capabilities of III-V semiconductor QDs for QKD are either limited to short-range fiber or free-space links[16] or rely on non-linear frequency conversion to shift the emission to a fiber-compatible wavelength.[15] Ultimately, QDs emitting at the wavelengths most compatible with optical fibers and atmospheric transmission windows[17] are essential for practical applications, such as quantum communication networks combining terrestrial optical fibers and satellite links.[18,19] This need has driven an active exploration and development of alternative QD materials[20,21] emitting in the 3rd telecom window using various epitaxy approaches including strain-driven Stranski-Krastanov (SK) growth[22] and droplet epitaxy[23] of InAs/InP QDs, and SK growth of InAs QDs on metamorphic layers grown on GaAs.[24,25,26] State-of-the-art demonstrations with these telecom-range QD systems include reports of multi-photon emission probability as low as $g^{(2)}(0) = (4.4 \pm 0.2) \times 10^{-4}$,[27] high Hong-Ou-Mandel (HOM) visibility of 98 % with post-selection[28] and 91.7 % without[29], as well as a high-rate intercity BB84 QKD demonstration with 10.9 kbits/s key rate and quantum bit error ratio of 0.65 %.[30]

We have recently introduced a new antimonide-based QD system exhibiting narrow-linewidth excitonic emission at 1470 nm[31], formed by GaSb-filling of droplet-etched nanoholes in AlGaSb. GaAs/AlGaAs QDs emitting at 780 nm grown by the same technique exhibit narrow exciton linewidths,[32] extremely small inhomogeneous broadening due to size uniformity,[33] bright single-photon emission,[34] and vanishing fine-structure splitting (FSS) owing to their dimensional symmetry and lack of strain-induced piezoelectric asymmetries.[35] These characteristics have enabled their use as a non-classical light source providing state-of-the-art performance in terms of photon indistinguishability and entanglement.[34,36,37,38] The newly developed antimonide-based QDs have already shown several beneficial features, such as extremely homogeneous ensemble emission,[31] a suitably low QD density necessary for single-QD devices,[31] and single-photon emission in the telecom S-band.[39] However, the advantages of the antimonide material system have not been fully exploited. For example, it offers wide band gap and strain tunability in addition to high refractive index contrast materials for developing efficient photonic structures like Bragg mirrors and waveguides. Moreover, they are compatible with monolithic integration on silicon by direct epitaxial growth because the lattice mismatch between GaSb-based materials and dissimilar substrates can be relaxed right at the first interface by the formation of a network of 90°-dislocations[40,41] and exploitation of nucleation layers.[42] These unique plastic properties have enabled the epitaxial growth of mid-infrared III-Sb semiconductor lasers on Si substrates.[43]

In a recently published work, we identified a 5 meV energy barrier for non-resonantly excited charge carriers to enter the GaSb QD from the surrounding medium.[44] While such a barrier can potentially protect the QD from charge fluctuations when employing coherent resonant excitation[32] or incoherent quasi-resonant excitation,[45] it introduces challenges in measuring QD properties when employing above-band excitation. The main challenges include:

- Slow emission decay characteristics and volcano-shaped autocorrelation response due to delayed feeding of charge carriers from a reservoir outside the QD;[39]
- Difficulty in studying the excitonic structure of a single QD due to highly convoluted spectral properties arising from simultaneously exciting a large number of charge complexes and multiple QDs;[31,39]

- Limited achievable photon count rate at excitation powers low enough to produce isolated exciton lines.

In this work, we introduce a single-photon emitter based on a single InGaSb/AlGaSb QD (Fig. 1a) with an indium content of 10 % corresponding to emission at 1500 nm wavelength region. A compact frequency-tunable single-frequency semiconductor laser operating in continuous-wave regime in the 1400–1470 nm range (Fig. 1b, Supplementary material Section 3) is used as the excitation source. This laser enables quasi-resonant pumping of the QD via its excited states and ground state via phonon-mediated pathways. Consequently, we gain access to the excitonic fine-structure of InGaSb/AlGaSb QDs and their single-photon emission characteristics without the convoluted dynamics caused by the charge reservoir.

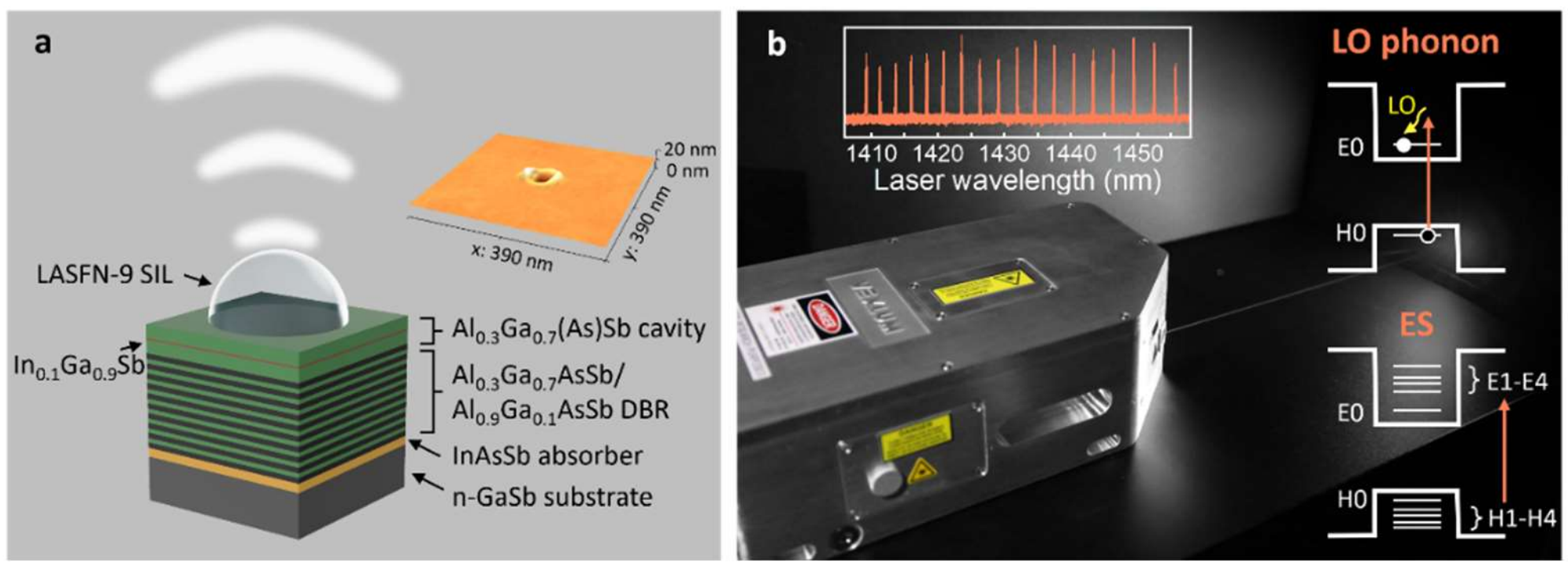

**Fig. 1: a** Schematic illustration of the antimonide-based QD device. The QDs, located in the middle of the AlGa(As)Sb λ-cavity, were grown by filling droplet-etched nanoholes with $In_{0.1}Ga_{0.9}Sb$. The surface profile of a nanohole is shown in the inset. The bottom mirror is a 9.5-pair AlGaAsSb distributed Bragg reflector (DBR). The device structure is separated from the substrate with an InAsSb absorber to block unwanted luminescence from the n-GaSb at the QD emission wavelength. A hemispherical solid immersion lens (SIL) is bonded on top of the semiconductor structure to enhance photon extraction. **b** The QD device is excited with a vertical external cavity surface emitting laser (VECSEL) providing single-frequency continuous-wave operation from 1400 to 1470 nm. These wavelengths allow LO phonon-assisted excitation of the QD ground state (H0-E0) and quasi-resonant excitation of transitions involving excited states (ES) of holes (H1-H4) and electrons (E1-E4).

## 2. Experimental details

### *2.1 Sample fabrication*

The epitaxial heterostructure shown in Fig. 1 was grown by molecular beam epitaxy (MBE). The layer sequence starts with a 400 nm thick lattice-matched InAsSb layer, which acts as an absorber to minimize unwanted photoluminescence (PL) emission from the n-GaSb(100) substrate. The absorber is then capped with a protective 2 nm GaSb layer after which a 9.5-pair bottom Bragg reflector mirror is grown with lattice-matched $Al_{0.9}Ga_{0.1}As_{0.07}Sb_{0.93}/Al_{0.3}Ga_{0.7}As_{0.03}Sb_{0.97}$ pairs. Finally, a 1λ-thick $Al_{0.3}Ga_{0.7}Sb$ microcavity is grown consisting of two outer ¼λ-thick $Al_{0.3}Ga_{0.7}As_{0.03}Sb_{0.97}$ and two inner ¼λ-thick As-free $Al_{0.3}Ga_{0.7}Sb$ pairs. The QDs are placed at the center of the cavity and are formed by Al-droplet mediated nanohole etching on the $Al_{0.3}Ga_{0.7}Sb$ surface and subsequent filling of the nanoholes by 7.2 monolayers (MLs) of $In_{0.1}Ga_{0.9}Sb$. The etching parameters follow those found in our earlier reports.[31,39] Namely, the droplets are etched at a temperature of 395 °C under a low Sb-flux of ~0.06 ML/s for 180 s. The geometry of the resulting nanoholes is shown in Fig. S1 in the supplementary material.

A hemispherical LASFN-9 solid immersion lens (SIL) was attached on top of the MBE-grown sample. The combination of a DBR back-reflector and SIL improves the photon collection efficiency but does not provide Purcell enhancement. The results of numerical finite difference time domain (FDTD) simulation of the collection efficiency and the far-field patterns of the QD emission are presented in Section 2 of the supplementary material. According to the FDTD simulations, introduction of DBR and SIL improves the photon extraction efficiency in a numerical aperture (NA) of 0.81 from 1.2 % to 16 %.

### *2.2 Optical experiments*

The optical experiments were carried out with a setup custom-built around a low-vibration pulse-tube cryostat cooled down to 5 K. A schematic of the complete setup is presented in Fig. S5 in the supplementary material. The sample was excited with the semiconductor laser described in section 2.3. An additional 950 nm LED was used as an optional above-band excitation source for two-color excitation. The power density of the LED was estimated as 200 $W/m^2$ at the back aperture of the objective and the area of the field of view. The LED intensity was optimized for charged trion (X*) emission intensity and kept at a level that would not produce detectable QD emission without the laser. The excitation beams were combined with non-polarizing beam splitters and focused on the sample with a 60-times magnifying NA = 0.81 objective located inside the cryostat. The laser spot size was around 1-2 μm. The same objective was used to collect the emitted photons. A 1470 nm long-pass filter was used to block the back-scattered laser beam during luminescence measurements. The polarization of the laser was controlled with a rotatable $\lambda/2$ waveplate. In polarization-dependent measurements of the QD emission, the detection polarization was filtered using a rotatable $\lambda/2$ waveplate and a fixed polarizer. The laser power was controlled with a variable neutral density filter. The rotation of the neutral density filter, waveplates, and the laser birefringent filter (BRF) was automated and integrated as a part of the spectral acquisition software.

The luminescence signal was dispersed on a thermoelectric-cooled InGaAs array with a 750 mm spectrograph equipped with a 600 l/mm grating. The resulting spectral resolution of the setup was 80 μeV. The single-photon counting was carried out using the secondary output port of the spectrometer. The photons corresponding to a single emission line were coupled into a 100 m long single-mode fiber to transfer them to another laboratory where the single-photon detection setup is located. The antibunching experiment was carried out at the other end of the fiber using a Hanbury-Brown-Twiss configuration consisting of a 50/50 fiber beam splitter and a pair of superconducting nanowire single-photon detectors. The photon statistics were counted using 39 ps bin size. The combined timing resolution of the detectors and the timing electronics is 40 ps.

### *2.3 Excitation laser*

The excitation laser used in this work is an optically pumped vertical-external-cavity surface-emitting laser (OP-VECSEL[46]). VECSELs combine the advantages of external-cavity solid-state disk lasers with those of quantum well (QW) semiconductor lasers and provide high output power, excellent beam quality, low noise, high spectral purity, and tunable wavelength versatile operation in a compact footprint. Although still perceived as an emerging laser technology, VECSELs have made considerable progress over the last decade[47,48] and are becoming key building blocks for Atomic, Molecular and Optical (AMO) physics experiments.

The VECSEL used in this work is a first-of-a-kind prototype of a VALO SF system from Vexlum Inc for QD excitation. The VALO SF laser platform has been particularly developed for atom and ion-based quantum systems,[47,48] offering low-noise operation down to sub-Hz linewidths,[49] wavelength flexibility, high beam quality, and accurate frequency tuning for tens of nm over longitudinal modes and GHz-level of hop-free tuning range, which is sufficient for scanning across the absorption line of a transform limited QD. The laser includes the VECSEL head, control electronics, and a low-noise chiller. The VECSEL head comprises a semiconductor gain mirror with AlGaInAs quantum wells with PL centered at 1428 nm at room temperature, an external cavity with 1 % output-coupling mirror mounted on a piezo actuator and with ~1 GHz free-spectral range (FSR), and a fiber-coupled 808 nm diode pump laser for carrier excitation. In a free-running configuration (i.e., with an empty cavity), the laser outputs >3.5 W of output power at around 1460 nm. To allow broadly tunable single-frequency operation between 1400–1470 nm, a 3-mm thick BRF is placed inside the laser cavity and further mounted on a rotation mount with resonant piezoelectric motors to enable precise remote control of the BRF rotation angle around its optical axis and thus the laser wavelength. During BRF rotation, a 6 kHz square wave with a 15 V amplitude is applied to the output coupler piezo to force the laser to jump through all cavity modes with 1 GHz resolution originating from the cavity FSR, which is fine enough to access all absorption lines of a QD. This driving allows fast hopping between cavity modes without skipping. The resulting wavelength tuning range is shown in Fig. S4a.

## 3. Results

The optical experiments were carried out for InGaSb QDs embedded in the device structure presented in Fig. 1a. The device combines a Bragg mirror back-reflector and a SIL for improving the photon collection efficiency into the available microscope objective. As explained in the supplementary material section 2, which deals with the optical design of the device, we expect around 16 % collection efficiency for NA = 0.81. First, we scan the BRF of the laser to sweep the lasing wavelength across the tuning range and record the back-reflected laser intensity as a function of wavelength. As shown in Fig. 2a, single-mode output is observed throughout the tuning range.

### *3.1 Photoluminescence excitation spectroscopy*

To identify absorption bands enabling quasi-resonant excitation of a single QD, we use photoluminescence excitation spectroscopy (PLE), where the QD emission spectrum in the 1500 nm range is recorded as a function of excitation wavelength. As shown in Fig. 2b, the intensity of the excitonic emission lines of the QD varies as a function of the excitation wavelength. Based on the PL spectrum of a QD ensemble (see Fig. S2 in the supplementary material), we expect the transition involving the first excited electron and hole states (E2-H2) to be at 1430 nm, which is at a separation of approximately 38 meV from the ground state transition energy. Since absorption can occur through transitions not visible in the PL spectrum, we use simulations performed within the combined framework of 8-band $kp$ and configuration-interaction methods (supplementary material section 8) to identify the peaks in Fig. 2b arising from transitions predominantly involving different electron and hole states. Fig. 2c. shows the simulated absorption spectrum, with the intensity of lines obtained from the transition oscillator strengths and their broadening accounting for both the natural Fourier-limited linewidth and an additional inhomogeneous broadening of 0.35 meV. The most pronounced peaks are related to transitions involving the lowest two hole states (H1 and H2) and the first excited electron states (E2 and E3). At the higher end of the laser tuning range, we observe a narrow resonance of the most pronounced emission line (X*) peaking at a detuning of 27 meV. This line arises from the excitation of the 1496.3 nm emission assisted by emission of a single LO phonon[50,51] (GaSb-like LO phonon in the $Al_{0.3}Ga_{0.7}Sb$ barrier[52]). In the following experiments, we excite the QD by targeting either the E1-H1 (neutral exciton) and E0H10 (negative trion) transitions at a detuning of 35 meV (BRF 30.4°) or the ground state E0-H0 through LO phonon-assisted excitation at a detuning of 27 meV (BRF 32.8°). From now on, we refer to optical pumping of carriers to the quasi-resonant E1-H1 and E0H10 transition as excited state (ES) excitation.

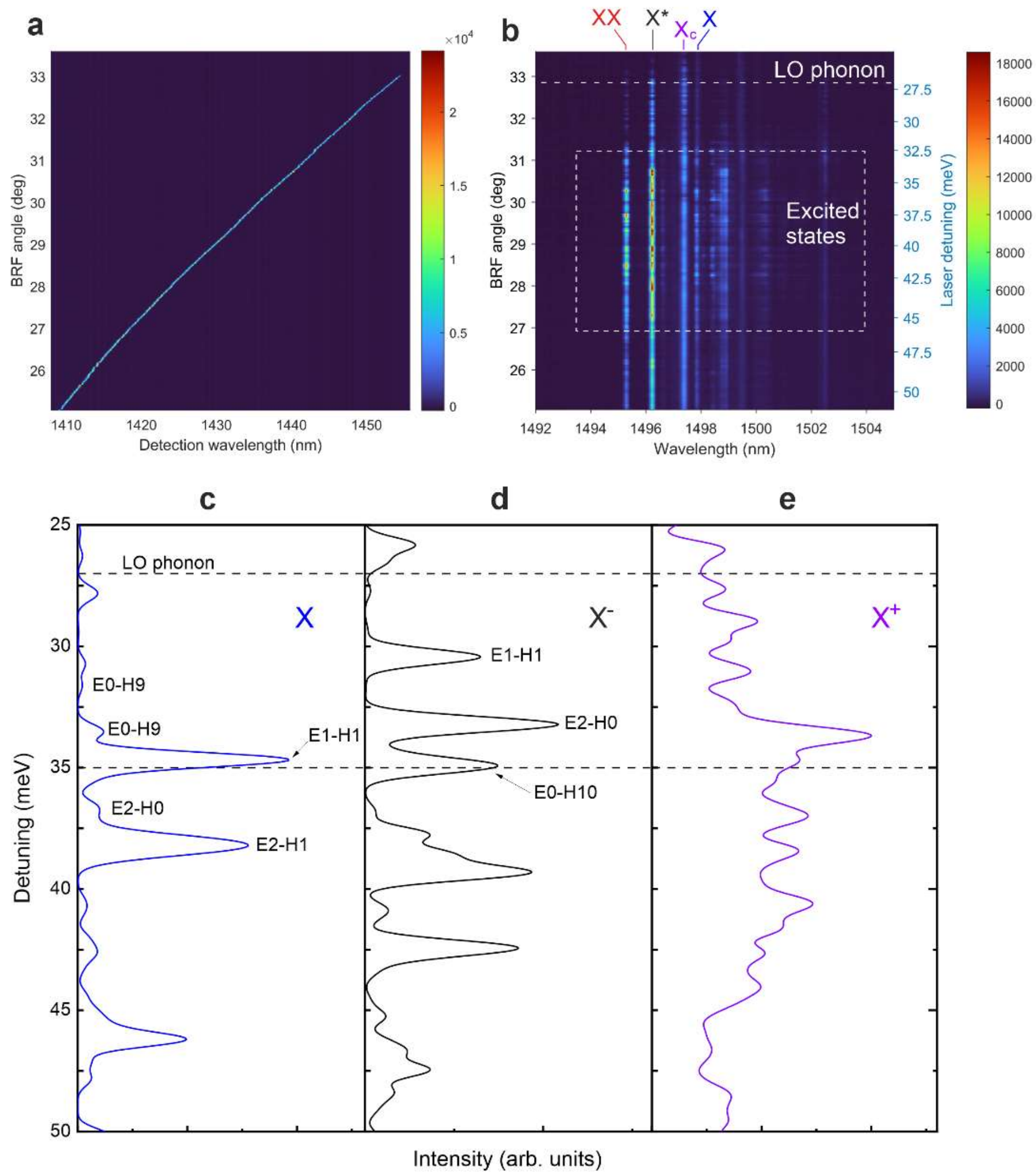

**Fig. 2: a** Single-mode lasing characteristic of the frequency-tunable excitation laser. The laser spectrum was detected from the intensity reflected from the QD device. The laser wavelength is tuned by rotating a BRF inside the laser cavity. **b** PLE results for a single InGaSb QD obtained with a laser excitation power of 4 μW maintained across different excitation wavelengths. Absorption spectra obtained from numerical simulations for **c** neutral exciton, **d** negative trion, and **e** positive trion. The dashed lines in **c**-**e** indicate the laser detunings used in the experiments for LO phonon-asssited excitation and ES excitation.

### *3.2 Excited state excitation*

In order to investigate the excitonic structure of the QD emission, we record the emitted PL spectrum as a function of detection polarization using ES excitation (Fig. 3). As shown in Fig. 3a, two emission lines exhibit fine structure evidenced by their polarization-dependent peak position. These lines are identified as a neutral exciton (X) at 1498.0 nm and biexciton (XX) at 1495.4 nm from power-dependent PL data (supplementary material section 7). The dominant emission line at 1496.3 nm showing no fine structure is identified as a trion (X*). Given the p-type nature of native defects in antimonide materials,[53,54] X* could be assigned as a positively charged trion based on the expected charge carrier statistics in the AlGaSb barrier and the intensity of the peak. However, the PLE response of the X* line (Fig. 3b) does not agree with the simulated absorption spectrum of positively charged trion (Fig. 3e) which is characterized by a large number of overlapping absorption peaks resulting in a quasi-continuous absorption band due to low separation of hole states. On the other hand, the simulated absorption spectrum of a negatively charged trion matches well with the PLE response of the X* line. Thus we assign X* line as a negatively charged trion. We attribute the predominant negative charging of the QD to trapping of photo-generated charge being in favor of lighter electrons over heavier holes due to the 5 meV energy barrier for charge transfer from QW to QD.[44] The weak periodic modulation of the X* line intensity with the polarization angle can be translated into a degree of linear polarization of 0.14 and further to a 1.5 % opposite-spin light hole admixture to the hole ground state.[55] The exciton line located between X and X* is then the positively charged trion, which shows no fine structure and has a power-dependency similar to the negatively charged X*. This line is referred to as $X_c$ in the supplementary material. We note also that the PLE response of the $X_c$ line is in agreement with the simulated absorption spectrum of a positively charged trion (Fig. e) resembling a quasi-continuous absorption band. The exciton peak assignment is supported by the excitation power dependent data shown in Fig. S8 in the supplementary material, where we observe a competition between the differently charged trions when the excitation power is changed. The positive charging due to the p-type carrier concentration of the barrier material dominates at low excitation powers, while a shift in favor of negative charging is observed when the concentration of photo-generated excess charge increases with increasing laser power. Noticeably, the dynamics can be shifted more in favor of the dominant negatively charged X* at 1496.3 nm by adding a 950 nm LED as a weak above-band excitation source (from now onwards referred to as two-color excitation[56]).

Now that we have provided an overview of the PL spectrum of a single InGaSb/AlGaSb QD and identified the different charge complexes, we can proceed to analyze the excitonic structure in more detail. A neutral biexciton binding energy, $\Delta E_{XX} = E_X - E_{XX} = -1.4$ meV (-2.6 nm), is obtained from the X-XX peak separation. Similar behavior is observed for another QD as shown in Fig. S7a in the supplementary material. The absolute values of $\Delta E_{XX}$ observed here are comparable to those seen in droplet-etched GaAs/AlGaAs QDs,[33,34,36,57] which, however, tend to have positive values. For the InGaSb/AlGaSb QDs investigated here, the observed negative values of $\Delta E_{XX}$ despite quite large confinement volume suggests some spatial separation of electrons and holes in a QD.[58]

An FSS of 24.1 ± 0.4 μeV is obtained from a sinusoidal fit to the XX peak position. Based on the investigation of 5 QDs, we conclude that this is a typical value for the investigated sample since we mainly observe FSS values in the 20-26 μeV range (Fig. S7. in the supplementary material), while an outlier with FSS = 5.1 ± 0.3 μeV is also identified. Overall, the observed FSS values indicate that further growth optimization for improving the QD symmetry is required to reduce the FSS to a level required for high fidelity of polarization entanglement. All investigated QDs have emission polarized in the same direction (Fig. S7c), which suggests that the FSS is mainly due to the QD shape asymmetry. As shown in Fig. S1, the lower part of the nanohole is highly symmetric while a clear asymmetry is observed towards the upper part near the ring structure around the nanohole. A possible route to reduce the FSS could be to reduce the in-filling amount and form a structure where the quantum confinement is predominantly defined by the symmetric lower part of the nanohole or symmetrizing the upper part of the nanohole and the surrounding ring structure. Reducing the intrinsic FSS values to 6-8 μeV in this manner would then make it possible to tune it down to zero by either strain-tuning[59] or AC stark tuning.[60]

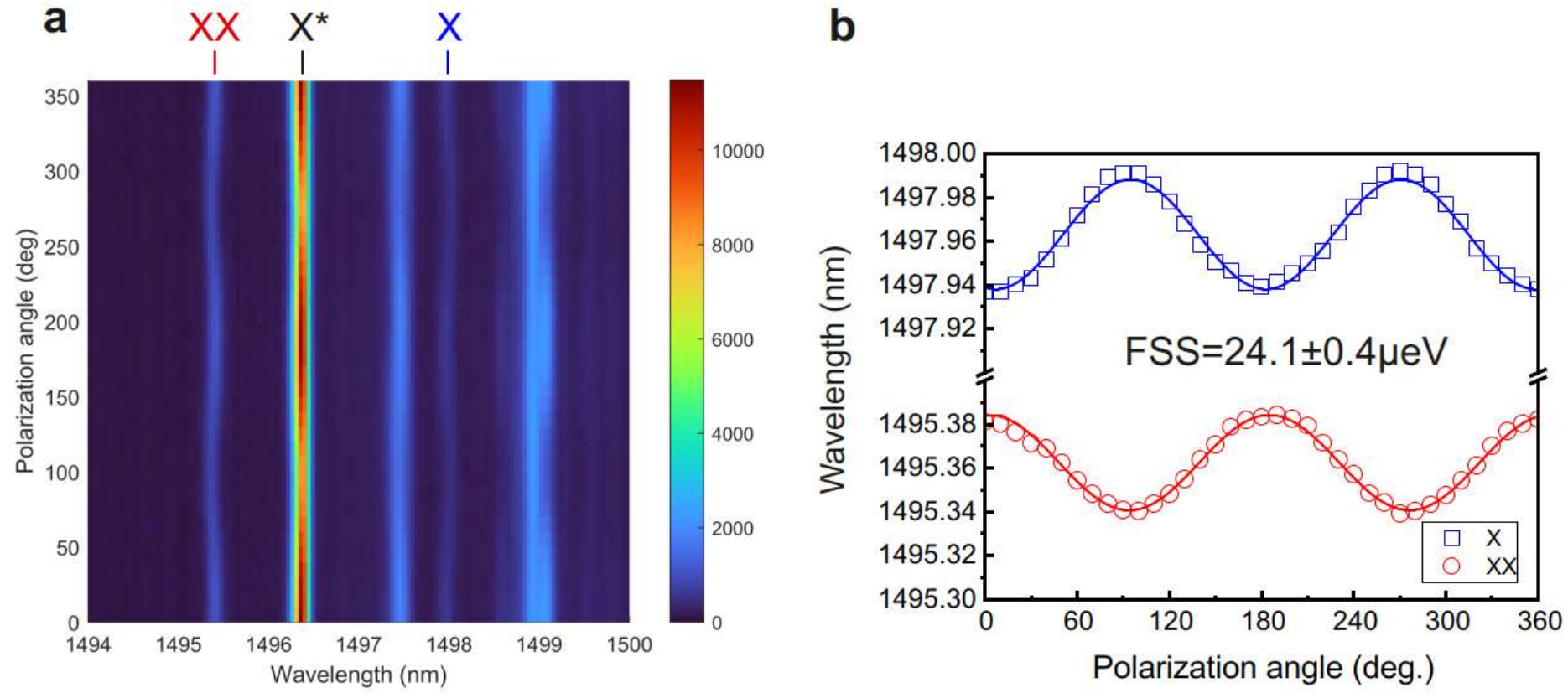

**Fig. 3: a** Polarization-resolved emission spectrum of a single QD with ES excitation. The X and XX emission lines exhibit FSS, with their emission wavelength varying periodically as a function of polarization, while the X* line is polarization-independent. **b** FSS evaluated from a sinusoidal fit to the peak wavelength values obtained from a Gaussian fit.

### *3.3 LO phonon-assisted excitation*

As shown in Fig. 3 and discussed above, the ES excitation scheme is effective for revealing the full excitonic structure and for investigating the fine structure of the confined states in the QD. It has been demonstrated for GaAs/AlGaAs QDs that it is possible to achieve similar $g^{(2)}(0)$ and HOM visibility using LO phonon-assisted excitation and strictly resonant fluorescence.[45] Therefore, we set the laser wavelength to 1449 nm for LO phonon-assisted excitation of the QD ground state. As shown in Fig. 4a, this results in an emission spectrum dominated by the X* transition. By implementing two-color excitation[61] (i.e. using an 950 nm LED for above-band excitation of the AlGaSb barrier in addition to the laser used for LO phonon-assisted excitation), we can further shift the charge dynamics in favor of X*, with 60 % of the spectrally integrated intensity now being emitted from the X* transition. The exciton linewidth is noted to be limited by the spectral resolution of the setup (80 μeV) both with and without two-color excitation.

Fig. 4b shows a second-order autocorrelation trace of the X* emission line counted in a fiber-based Hanbury-Brown-Twiss configuration (40 ps timing resolution) using the two-color excitation scheme. The experimental data is fitted with a model that includes photon antibunching (single-photon emission) and spectral blinking.[62] The model parameters include the multi-photon emission probability $g^{(2)}(0)$, the emission on-time $\beta$ related to blinking process, as well as the blinking time constant $\tau_{Blink}$ and the correlation time $\tau_{Corr}$, which includes the radiative lifetime of the exciton, and in case of incoherent excitation, additional contributions from relaxation and capture dynamics. From the fit we obtain a $g^{(2)}(0) = 0.05 \pm 0.03$, which compares well with the value of $0.019 \pm 0.001$ reported for GaAs/AlGaAs QDs under similar experimental conditions (LO phonon-assisted excitation)[50] but does not reach the record values of telecom QDs[27,28]. We expect to observe much lower values of $g^{(2)}(0)$ in the future for our InGaSb/AlGaSb QDs by (i) subduing emission from the GaSb substrate which currently contributes detrimentally to the autocorrelation experiments and (ii) improving the low count rate due to the inefficient fiber coupling which ultimately affects the $g^{(2)}(0)$ fitting process.

The observed blinking characteristics are similar to the values reported in GaAs/AlGaAs QDs without applying an electric field bias to stabilize the charge environment. The emission on-time, $\beta = 0.35 \pm 0.007$, is similar to what is seen in GaAs/AlGaAs QDs under resonant excitation[63] and reasonably better than GaAs/AlGaAs QDs under quasi-resonant excitation,[62] while $\tau_{Blink} = 26.2 \pm 1.1$ ns is also similar to the GaAs/AlGaAs QDs.[62] Fig. 4c shows the second-order autocorrelation trace of the X* emission line under ES excitation. The result from the fit is quite similar to the LO phonon-assisted excitation for $\tau_{Blink}$, but a slightly larger emission on-time with $\beta = 0.44 \pm 0.01$ and $g^{(2)}(0) = 0.01 \pm 0.05$ are revealed. It should be noted that while the autocorrelation characteristics are quite similar for the two excitation methods, LO phonon-assisted excitation provides a higher count rate thanks to the emission predominantly by X* recombination. This is particularly so when the two-color excitation is applied.

The correlation time $\tau_{Corr}$ defines the width of the antibunching dip at zero-delay. In the simplest case of a two-level system, it would be determined by the spontaneous radiative lifetime of the emitter. However, in the case of quasi-resonant excitation, the value of $\tau_{Corr}$ may additionally include the component of temporal delay arising from the thermalization process involved in the spontaneous radiative decay from the X* state. This is particularly the case for the ES excitation, which is better described as a three-level system. For LO phonon-assisted excitation, we get $\tau_{Corr} = 0.84 \pm 0.05$ ns, while for ES excitation, it is slightly slower with $\tau_{Corr} = 0.98 \pm 0.08$ ns. These values are somewhat larger than 0.32 ns reported for weakly confined GaAs/AlGaAs QDs,[62] but of a similar magnitude to typical radiative lifetimes of radiatively efficient systems like In(Ga)As/GaAs QDs emitting at 920 nm[64] and InAs/InGaAs QDs emitting at 1550 nm.[26] The true radiative lifetimes of the InGaSb/AlGaSb QDs will be assessed in our future work using a pulsed semiconductor laser, which is currently under development.

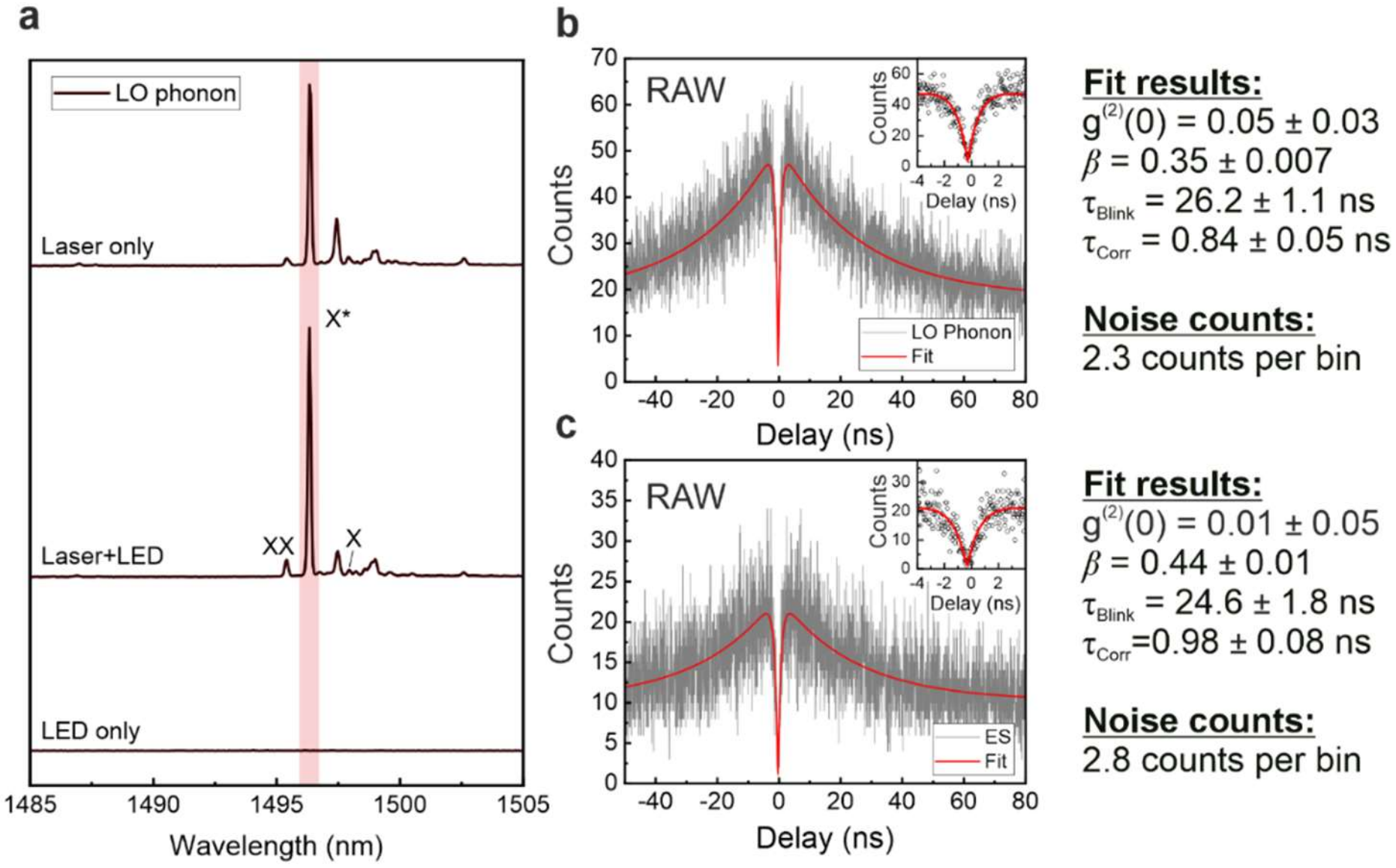

**Fig. 4: a** Single-QD emission spectrum obtained with 4.5 μW LO phonon-assisted excitation. The plot shows the spectrum for excitation with laser only and for two-color excitation, where a 950 nm LED was added for tuning the charge distribution in the AlGaSb barrier. The LED power density was set to 200 W/m$^2$, which was low enough not to produce detectable emission when the laser was switched off. **b** Second-order autocorrelation trace of the X* line with LO phonon-assisted two-color excitation.[56] **c** Second-order autocorrelation trace of the X* line with ES excitation. A two-level model was fitted to the autocorrelation traces to obtain the physical parameters of quantifying photon antibunching and spectral blinking.[62] The insets in **b** and **c** show a zoomed-in view of the zero-delay region.

## 4. Discussion

We have introduced InGaSb/AlGaSb QDs formed by filling droplet-etched nanoholes as a new single-photon source emitting in the 1500 nm wavelength range. A compact frequency-tunable semiconductor laser is utilized for PLE spectroscopy, which is carried out within the excitation wavelength range, covering both quasi-resonant ES excitation and LO phonon-assisted excitation of the QD ground state. By exciting the QD states directly, we can eliminate the complex charge dynamics arising from delayed feeding of charge carriers from a reservoir outside the QD, which are present when using above-band excitation. Consequently, we achieve spectrally pure emission predominantly from a single charged exciton line (X*) and pure single-photon emission with $g^{(2)}(0) = 0.05 \pm 0.03$. This is a considerable improvement compared to our previous work with above-band excitation, resulting in non-ideal spectral features and $g^{(2)}(0) = 0.16 \pm 0.02$.[39] These results compare well with the data collected from GaAs/AlGaAs QDs under similar experimental conditions in terms of single-photon purity, spectral features, and blinking characteristics.[50,62] The InGaSb/AlGaSb QDs are thus expected to respond well to the application of an electric field bias from a p-i-n diode for charge stabilization, which is known to completely eliminate blinking in GaAs/AlGaAs QDs.[32,63]

Furthermore, exciting the QD states directly enabled us to obtain a detailed overview of the excitonic structure. Additionally, this allowed for measuring the key physical properties relevant to further developing antimonide QDs for high-quality single- and entangled-photon emission in the third telecom window. In particular, the biexciton binding energy $\Delta E_{XX}$ = -1.4 meV (-2.6 nm) is large enough to render possible coherent pumping of the biexciton-exciton cascade by two-photon absorption.[32,63] Moreover, the FSS values obtained from the investigated QDs ranges from 5 to 26 μeV (Fig. 3, Fig. S7), with the larger end of the range being more representative of the population. While these values are not small enough for the generation of polarization-entangled photons with high fidelity, they are a good starting point for optimization. Since the nanoholes etched in AlGaSb with Al droplets are highly symmetric (Fig. S1 in the supplementary material), the FSS can result from asymmetric filling, piezoelectric asymmetry resulting from the 0.63 % lattice-mismatch between the $In_{0.1}Ga_{0.9}Sb$ QD material and the GaSb substrate, atomistic details of QD material alloying, or heavy-light hole mixing enhanced due to low biaxial strain even in an almost symmetric QD. It should be noted, however, that the expected lattice strain is an order of magnitude smaller than, for example, in the InAs/GaAs system (7 %). Nevertheless, there is viability for reducing the indium content and simultaneously increasing the QD size to reduce the alloying and piezoelectric effects and increasing the heavy-light hole splitting, while still maintaining emission in the 1500 nm wavelength region. Such a change in the QD structure would shift the resulting energy structure towards even weaker confinement and provide certain benefits such as positive biexciton binding energy and a faster radiative rate as seen in GaAs/AlGaAs QDs.[45] It has been recently shown that high-indium-content $In_xGa_{1-x}As$/AlGaAs QDs (with x=0.1-0.4) grown in droplet-etched nanoholes can exhibit as low FSS as pure GaAs QDs with no clear correlation between FSS and indium content.[65] Therefore, it is more likely that cause for the FSS observed in our $In_{0.1}Ga_{0.9}Sb$/AlGaSb is related to QD shape asymmetry. To this end, symmetrizing the upper part of the nanohole and the surrounding ring structure by careful adjustment of the epitaxial growth has played an important role in minimizing the FSS in GaAs/AlGaAs QDs.[66] This approach can be adopted to the our InGaSb/AlGaSb QDs.

## 5. Conclusion

In summary, we revealed the fundamental optical properties of InGaSb/AlGaSb QDs formed by filling droplet-etched nanoholes. The QDs are embedded in a device structure comprising an antimonide-based Bragg reflector and a solid immersion lens to improve photon collection. Exciting the InGaSb/AlGaSb QDs quasi-resonantly with a compact frequency-tunable semiconductor laser allowed us to study their excitonic fine structure with unprecedented detail. Furthermore, directly addressing the energy states of single QDs allowed us to achieve predominant emission from a single exciton line with a low multi-photon emission probability of $g^{(2)}(0) = 0.05 \pm 0.03$. These results represent a major step towards developing antimonide-based QD materials for quantum photonics and applying a wavelength-flexible semiconductor laser platform as an excitation source for deterministic quantum emitters in secure fiber-based quantum networks.

**Funding.** Finnish Research Council project QuantSi (Decision No. 323989); Finnish Research Council project CryoLight (Decision No. 357351); *Business Finland co-innovation project QuTI (41739/31/2020); FiGAnti project (Grant Agreement No. 101017733) funded within the QuantERA II Programme; Business Finland Rise to Challenge project TeleQuant (1835/31/2025); National Science Centre, Poland - project 2023/05/Y/ST3/00125.*

**Acknowledgement.** Financial support from the Finnish Research Council projects QuantSi (Decision No. 323989) and CryoLight (Decision No. 357351), Business Finland co-innovation project QuTI (41739/31/2020), Business Finland Rise to Challenge project TeleQuant (1835/31/2025), the Research Council of Finland Flagship Programm PREIN, and FiGAnti project funded within the QuantERA II Programme that has received funding from the European Union's Horizon 2020 research and innovation program under Grant Agreement No. 101017733 and National Science Centre, Poland - project 2023/05/Y/ST3/00125 is acknowledged. TH acknowledges financial support from Tampere University Institute of Advanced Study. M. G. and R. M. are grateful to Krzysztof Gawarecki for sharing his computational code. The authors would like to thank Prof. Robert Fickler, Lea Kopf, and Jaime Moreno Zuleta for their help in the single-photon autocorrelation experiments.

**Disclosures.** Authors declare no conflicts of interest.

**Supplementary material.** Quantum dot and device characterization; Reflector design and extraction efficiency; Laser characterization; Experimental setup; Additional data for polarization-dependency; Power-dependency of excitons; Theoretical modeling of quantum dot properties.